\documentclass[aps,pre,twocolumn,groupedaddress,showpacs]{revtex4}
\usepackage{amssymb}
\usepackage{graphicx,color}
\definecolor{r}{rgb}{1,0,0}
\definecolor{g}{rgb}{0,1,0}
\definecolor{b}{rgb}{0,0,1}

\begin{document}


\title{Rain water transport and storage in a model sandy soil with hydrogel particle additives }


\author{Y. Wei$^{1,2}$, D. J. Durian$^1$ }
\affiliation{$^{1}$Department of Physics and Astronomy, University of Pennsylvania, Philadelphia, PA 19104-6396, USA}
\affiliation{$^{2}$Complex Assemblies of Soft Matter, CNRS-Rhodia-UPenn UMI 3254, Bristol, PA 19007-3624, USA}


\date{\today}

\begin{abstract}
We study rain water infiltration and drainage in a dry model sandy soil with superabsorbent hydrogel particle additives by measuring the mass of retained water for non-ponding rainfall using a self-built $3$D laboratory set-up.  In the pure model sandy soil, the retained water curve measurements indicate that instead of a stable horizontal wetting front that grows downward uniformly, a narrow fingered flow forms under the top layer of water-saturated soil.  This rain water channelization phenomenon not only further reduces the available rain water in the plant root zone, but also affects the efficiency of soil additives, such as superabsorbent hydrogel particles.  Our studies show that the shape of the retained water curve for a soil packing with hydrogel particle additives strongly depends on the location and the concentration of the hydrogel particles in the model sandy soil.  By carefully choosing the particle size and distribution methods, we may use the swollen hydrogel particles to modify the soil pore structure, to clog or extend the water channels in sandy soils, or to build water reservoirs in the plant root zone.

\end{abstract}

\pacs{to be determined}
%


\maketitle




Sandy soils are widely distributed in the world and are known as the hungry soil due to their poor water-holding capacity and low nutrients.  To grow plants in sandy soils, additives such as superabsorbent hydrogel particles~\cite{Buchholz, Kazanskii92, Bouranis95} have been developed to improve the soil water retention and to prevent nutrient runoff.  The study of hydrogel particle additives started in the early 1980's~\cite{Azzam80, Azzam83, Flannery82, Johnson84}.  To date, such additives have been proven to significantly increase the soil moisture content~\cite{Flannery82, Johnson84, Singh, Bhardwaj07, Abedi08, Bai10} and affect the water transport in sandy soils~\cite{Elshafei92, Elshafei94, Bhardwaj07, Abedi08, Verneuil11, WeiPressurePlate}.  However, most of these studies were conducted using a complex plant-soil system with varied irrigation conditions.  A few studies that used simplified model sandy soils either focus on the effect of swollen hydrogel particles in fully-saturated environment such as hydraulic conductivity~\cite{Verneuil11} and soil water retention curve measurements~\cite{WeiPressurePlate}, or emphasize visualizing the swelling of hydrogel particles under sufficient water supply such as 2-dimensional water infiltration studies~\cite{Singh}.

The situation in the field may significantly differ from the assumptions made in those experiments, since rain and irrigation water creates preferential paths when infiltrating into sandy soils.  Both field evidence~\cite{Vanommen89, Ritsema93, Hendrickx93, Ritsema98, Williams00} and laboratory researches~\cite{Selker92, Babel95, Yao96, Tullis07, Annaka10} have shown that fingered flows form in dry sandy soils under non-ponding rainfall.  This water channelization phenomenon may be caused by the structure heterogeneity of soils or the space inhomogeneity of water supply, but also may be induced in uniform sandy soils by instability of a downward-moving wetting front.  On one hand, the efficiency of the soil additives may not be correctly determined without accounting for this phenomenon.  On the other hand, understanding the effects of hydrogel particle additives on rain water channelization may help us to further improve the soil additives, optimizing both their physical properties and their distribution methods.

In this paper, we build a 3-dimensional laboratory set-up that measures the mass of the retained water in a model sandy soil under controlled rainfall.  The model sandy soil has a well-known pore structure and surface wetting properties.  The water flow profile inside the soil packing is determined based on the curve of excess retained water.  Then a commercial product of hydrogel particles with a carefully chosen size are added into the model sandy soil with different distribution methods, for example uniformly mixed or placed in a layer.  The retained water curves during and after rainfall are measured so that the effects of hydrogel particle additives on water flow profile in the model sandy soil can be determined and discussed.


\section{Experiment}

\begin{figure}
\includegraphics[width=3in]{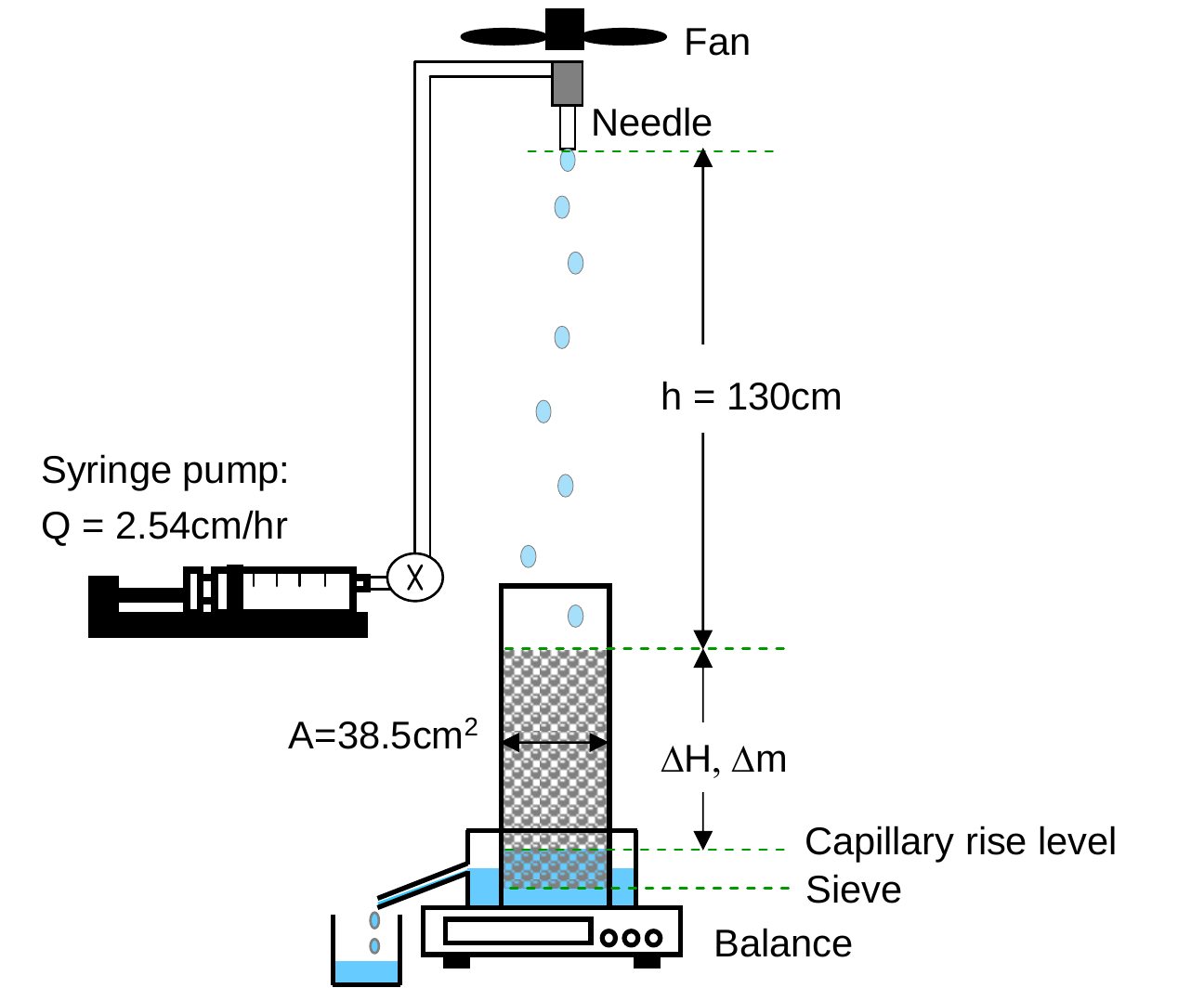}
\caption{(Color online)  Schematic of the laboratory set-up for the retained water measurements.  A needle is connected to a syringe pump and suspended under a small fan.  The syringe pump provides a fixed flow rate of $Q$.  Fan makes rain droplets created on the needle tip randomly impinging on the surface of a soil packing.  The soil packing has a cross-sectional area of $A=38.5$~cm$^2$ and is placed in a drainage container with water in it.  Due to the capillary rise effect, a water table is then formed on the bottom of the soil packing. The capillary rise level maintains the same during and after rainfall.  $\Delta H$ here represents the height of a soil packing above the water table and $\Delta m$ represents the mass of the retained water in that region.  Before rain starts, the mass of the retained water equals zero.  During and after rainfall, the mass of the retained water is recorded as a function of time. }
\label{Setup}
\end{figure}

Fig.~\ref{Setup} shows a schematic of the laboratory set-up.  A polypropylene dispensing needle (Intramedic, BD Inc.) is connected to a syringe pump (Harvard Apparatus Inc.) by Tygon tube and then fixed on the front grill of a $6$~inches clip-on fan (Air King Inc.).  The fan is face down and clipped on a stage about $1.5$~m higher than the bench.  The syringe pump infuses deionized (DI) water to the needle at a fixed volumetric flow rate $Q$ to create rain droplets.  The size and the frequency of the droplets can be controlled by modifying the needle inner diameter and the pump infusing rate.  The fan creates turbulence to perturb the horizontal impinging location of the rain droplets, and slightly accelerates the droplets vertically.  By adjusting the speed of the fan, we ensure a random impingement of the rain droplets on the surface of a soil sample packing placed under it.

The sample column is made by fitting a $30$~cm height cylindrical Plexiglass tube into a standard copper sieve (Dual Manufacturing Co.).  The tube has hydrophobic surfaces and a inner cross-sectional area of $A=38.5$~cm$^2$. The mesh size of the copper sieve is $0.106$~mm; our tests show that it doesn't have a detectable influence on the water flows.  The sample column is placed in a home-made drainage container with water in it, kept at a fixed height by a drainage outlet.  An electrical balance (Citizen Inc.) with a resolution of $0.1$ gram is used to measure their mass change ($\Delta m$) during and after rainfall.   Due to capillarity, the water level in a soil packing will be higher than that in the drainage container and this wetted saturation zone mimics the water table in the real field.  A key advantage of our set-up is that it allows for the possibility of a steady-state situation, where the rate of rain onto the sample equals the flow rate of water from the drainage outlet -- in which case the configuration of the liquid channels  within the model soil would be constant.

\subsection{Materials}

The model sandy soil we used is a random close packing of glass beads (Potters Industries Inc.) with a diameter of $1$~mm ($\pm20$\%).  For cleaning, they were first burnt in a furnace at $500^{\circ}$C for 72~hours and then soaked in a $1$M HCl bath for one hour.  After that the glass beads were rinsed with plenty of DI water, baked in a vacuum oven at about $110^{\circ}$C for 12~hours, and cooled to room temperature.  The clean beads have hydrophilic surfaces.  By weighting several hundreds of beads together, and counting the actual number carefully, we determine that the average mass of a single glass bead is $m_{bead}=1.17\pm0.01$~mg.

The hydrogel particle additives are a commercial product provided by Degussa Inc.(Stockosorb SW).  They are $50/50$ potassium acrylate and acrylamide cross-linked together through industrial polymerization.  Since the particles are obtained by grinding hydrogel bulks, their shapes are faceted and random.  The hydrogel particles we used are further sieved to be between $0.3$~mm and $0.5$~mm sized meshes.  The average mass of a single hydrogel particle is $m_{gel}=0.067\pm0.002$~mg, obtained by counting over $500$ dry particles and weighting them together.  To determine the global water holding capacity of these particles, we place $0.01$~gram dry particles in plenty of DI water.  After three hours, the swollen particles are collected and wiped.  Their swollen mass is around $300$ times higher than that in dry.  We also check the swelling ratio $\delta$ in the long axis of hydrogel particles using a microscope and find they swell to $6$ times their dry length when placed in contact with DI water for three hours.  These values are used in the determination of the gel layer number.  When exposed in atmosphere at room temperature, the swollen hydrogel particles de-swell slowly so as to lose $85$\% of their stored water within a day.  The cycle of swelling and de-swelling repeats again and again without degradation when their surrounding environment changes, and the ability of storing and releasing water makes them a good candidate in sandy soil moisture modification.

\subsection{Procedures}

We first set the raining condition.  According to literature~\cite{Byers59, Rogers79, Hobbs81, Pruppacher10}, the diameter $d$ of rain droplets in a real rain varies from $0.5$~mm to $4$~mm depending on the rain rate, and the corresponding terminal falling speed $U_t$ varies from $2.1$~m/s to $8.8$~m/s.  Therefore, we choose a rain rate of $Q=2.54$~cm/hr for a heavy rain.  A high speed camera (Phantom $630$, Vision Research Inc.) is used to record the motion of the droplets right before they impact the soil surface.  We determine that the diameter of the rain droplets is $d=3.0\pm0.1$~mm and the impact speed of the rain droplets is $U_t=6.1\pm0.1$~m/s for a falling height of $h=130$~cm.  Due to the influence of the fan, this value of $U_t$ is about $1.0$~m/s higher than that estimated from a free fall.

To prepared a soil packing, we carefully pour dry glass beads into the sample column, $1$ to $2$~cm each time, to the designed packing height.  The sample column is gently tapped from time to time to get a random close packing.  The tetrahedral holes in such a packing have a radius of $2R_{tet}=0.225$~mm, which is a bit smaller than the axis of the dry hydrogel particles.  Therefore, when hydrogel particles are mixed in, either uniformly or placed in a layer at certain depth, they will not be moved or carried away by the infiltrating flow of rain water.

The partially-filled sample column is then placed in the dry drainage container, right under the rain source.  Water is slowly added into the container until the designed drainage level is reached and excess water begins to drain out.  The designed drainage level is about $1.5$~cm higher than the mesh position inside the sample column.  Since water can freely flow in or out of the sample column through the mesh, a horizontal wetting front appear inside the soil packing and moves upward due to capillarity~\cite{Delker96}.  We keep adding water until it rises to a stable height.  The final capillary rise height in the model soil packing is $1.0\pm0.1$~cm.  A cover is then added on the drainage container to minimize the evaporation in the system.  During and after rainfall, the position of the wetting front is monitored to make sure that it is stable all times.

We start a rain at $t=0$ and record the balance reading once per minute.  The mass of the retained rain water in a soil packing is then determined as
\begin{equation}
   \Delta m = m_t - m_0,
\label{Dm}
\end{equation}
where $m_0$ is the reading at $t=0$ and $m_t$ is the reading at time $t$.  We stop the rain when the reading is quasi-stable or when the soil packing is fully saturated to top, and call that time as $t_{stop}$.  The mass of the drainage water after rain stops is then given as
\begin{equation}
   \Delta m - \Delta m_{stop}= m_t - m_{stop},
\label{DmStop}
\end{equation}
where $m_{stop}$ is the reading at $t=t_{stop}$ and $\Delta m_{stop}$ is the mass of the retained water at $t=t_{stop}$.  The mass of the retained water is then scaled by water density $\rho$ and the sample cross-sectional area $A$ so that it has units of length, to better correspond with rain rate in the usual units of length per time.


\section{Pure model sandy soil}

\begin{figure}
\includegraphics[width=3in]{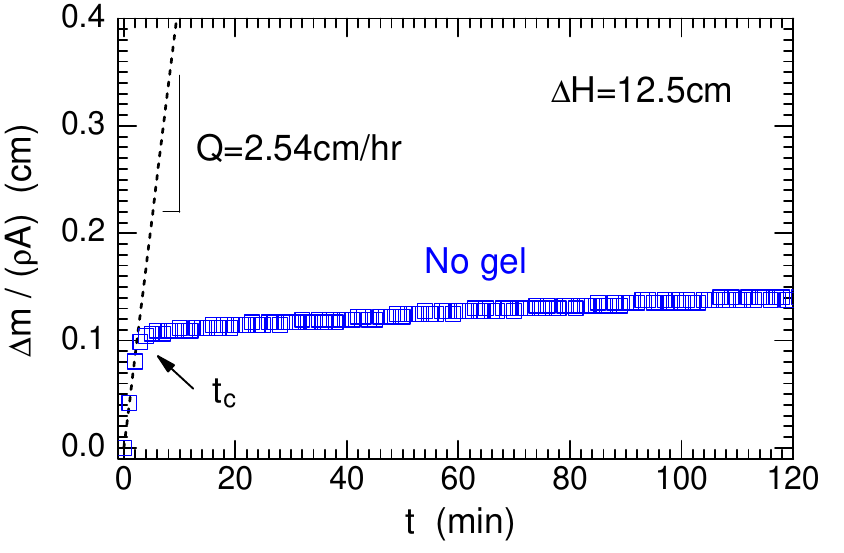}
\caption{(Color online)  Variation of the retained water with time for a model sandy soil, $1$~mm hydrophilic glass beads, at a rain rate of $Q=2.54$~cm/hr.  The mass of the retained water $\Delta m$ is scaled by water density $\rho$ and the cross-sectional area $A$ of the soil packing.  Rain starts at time $t=0$.  The dry glass beads above the water table have a packing height of $\Delta H=12.5$~cm.  During the rain, the water table in the soil packing is stable.  In the first few minutes of rain, the retained water increases the same as the rain rate.  After that it slowly approaches a stable value.  $t_c$ marks the time at which the water channel reaches the water table and fully penetrates the whole dry packing.  }
\label{NoGel}
\end{figure}

\begin{figure}
\includegraphics[width=3in]{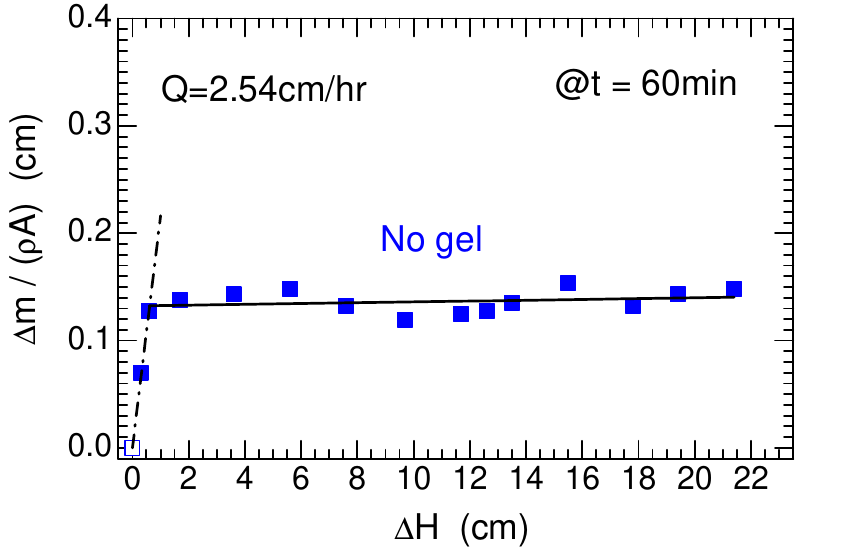}
\caption{(Color online)  Variation of the retained water after an hour of rain ($t=60$~min) at a rain rate of $Q=2.54$~cm/hr with the packing height $\Delta H$ above the water table for a model sandy soil, $1$~mm hydrophilic glass beads.  The mass of the retained water $\Delta m$ is scaled by water density $\rho$ and the cross-sectional area $A$ of the soil packing.  Solid line is a linear fit for $\Delta H \geq 0.6$~cm data.  The slope of the fit confirms two things: the water channel is very narrow compared to the packing size, and the channel is in partially saturated state during the rain.  }
\label{NoGelDH}
\end{figure}

We begin by first determining retained water curves in the pure model sand soil of glass beads, with no hydrogel additives.  Fig.~\ref{NoGel} shows an example of the mass of retained water as a function of time.  $\Delta H=12.5$~cm labeled in the figure refers to the height of the soil packing above the fully-saturated water table, which exits before the rain starts.  We see that at first the retained water increases the same as the rain rate, $Q$.  But after a few minutes, at a time we denote $t_c$, it abruptly deviates and thereafter increases only very slowly.  The quasi-stable mass retained water, around $\Delta m/(\rho A)\approx0.12$, is less than one tenth of the water  required to fully saturate the whole dry region of the soil packing.  With this same procedure, we then vary the packing height $\Delta H$ from $0$ to $22$~cm and repeat the same measurement over ten times.  Similar water retention curves are obtained for these packings, but sometimes an overshoot may occur in the transition point of those curves.  Fig.~\ref{NoGelDH} summarizes the quasi-stable values of the retained water in those soil packings at different heights, which we estimate based on the behavior at $t=60$~min.  The results are close to each others except those with packing height less than $0.6$~cm.

Based on Fig.~\ref{NoGelDH} and visual observations, we come to the conclusion that rain water only uniformly infiltrates into a shallow top layer of the model sandy soil and after that it creates a narrow channel penetrating down through the whole dry region of the soil packing.  Once the water channel connects to the water table in the soil packing, any additional rain water flows down and out the channel and the system reaches a quasi-stable state.  We speculate that the subsequent observed slow increase in retained water mass is not due to a gradual thickening of the channel or the top layer, but rather due to adsorption of water vapor as a wetting layer on the glass beads (and in the hydrogel particles, in the next sections).

The two lines fit to the data in Fig.~\ref{NoGelDH} support this picture.  For $\Delta H$ smaller than the thickness of the wetted shallow top layer, the mass of retained water should be proportional to $\Delta H$ consistent with the first line.  The slope has a value of $0.22\pm0.01$, which gives the volume fraction of water in the top layer.  If it were fully saturated, the value would be 0.36 since the packing fraction of random-packed spheres is 0.64.  Thus, the wetted top layer is a little more than half saturated.  This behavior persists until about $\Delta H=1$~cm, which corresponds well with visual observation of the thickness of the wetted top layer.  For taller packings, a water channel forms and the mass of retained water increases with $\Delta H$ due only to the increase of water volume in the channel.  In particular, the slope should be set by $d \Delta m/d \Delta H = \rho A_w$ where $A_w$ is the cross-sectional area of the {\it water} in the channel.  The slope of the second line in Fig.~\ref{NoGelDH} is $(3.8\pm3)\times 10^{-4}$, which gives $A_w=(3.8\times 10^{-4})A=0.015$~cm$^2$.  This compares to the size of the tetrahedral hole between glass beads as $A_w/(\pi R_{tet}^2)=30$.  Thus the radius of  the actual water channel is about $\sqrt{30/\pi}=3$ beads.  This is consistent with observation made by turning off the rain and quickly excavating the sample.  We note that actual channels do not have a perfectly constant cross section, nor are they perfectly vertical.

With these results as a baseline, we are now ready to modify the pure model soil by addition of hydrogel particles and measure their effects on the retained water curve.


\section{Hydrogel particles mixed in uniformly}

One controlled way to apply hydrogel particle additives is to uniformly mix them with the glass beads.  We first use this distribution method to study the effects of hydrogel particle concentration and mixing depth.  A model sandy soil packing with packing height $\Delta H=12.5$~cm above the water table is used as the base packing (`No gel' packing).  Different amount of hydrogel particles are uniformly mixed into a shallow or a deep layer at the top of the dry soil packing.  We determined the gel number ratio $\alpha$ in a mixture region as
\begin{eqnarray}
   \alpha &=& N_{gel} : N_{bead}  \\
          &=& \frac{M_{gel}/m_{gel}}{M_{bead}/m_{bead}} ~.
\label{GelNumRatio}
\end{eqnarray}
Here $N_{gel}$ and $N_{bead}$ are the total number of the added hydrogel particles and the total number of the glass beads in the mixed region; $M_{gel}$ and $M_{bead}$ are the total mass of the added hydrogel particles and the glass beads in the mixture region respectively; $m_{gel}$ is the average mass of a single gel particle while $m_{bead}$ is the average mass of a glass bead.

\begin{figure}
\includegraphics[width=3in]{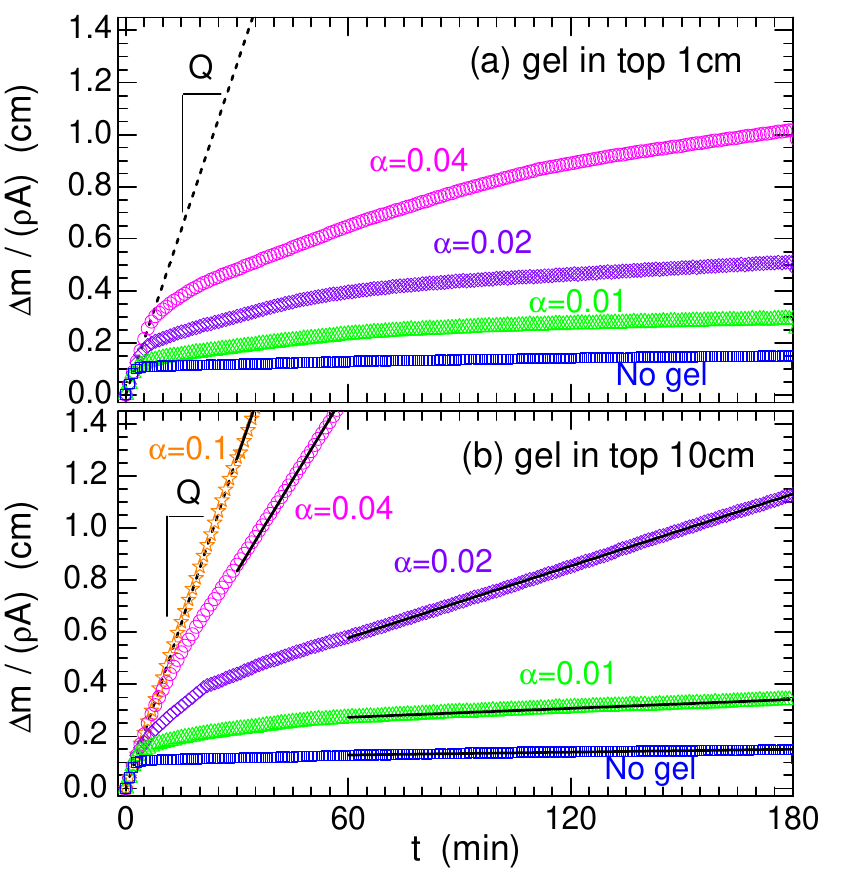}
\caption{(Color online)  Variation of the retained water with time for $1$~mm hydrophilic glass bead packing under a heavy rain, with dry hydrogel particles ($0.3-0.5$~mm in axis) uniformly mixed into (a) top $1$~cm region and (b) top $10$~cm region of the soil packing.  The initial packing height above the water table is $\Delta H=12.5$~cm for all packing.  $\alpha$ is the number ratio of hydrogel particles to glass beads in the mixture region and is determined in Eq.~(\ref{GelNumRatio}).  The mass of the retained water $\Delta m$ is scaled by the water density $\rho$ and the cross-sectional area $A$ of the packing.  Rain starts at time $t=0$ with a rain rate of $Q=2.54$~cm/hr.  During the rain, the water table remains stable.  Dashed line marks rain water; the solid lines are linear fits to determine the cumulate rate of rain water in each packing.  As $\alpha$ increases, more rain water is retained in the packing.  For a sufficient high $\alpha$, the retained water curve overlaps with the dashed line of the rain rate.  }
 \label{Mixture}
\end{figure}

Fig.~\ref{Mixture}~(a) shows the retained water curves for packings with dry hydrogel particles uniformly mixed into top $1$~cm region under a heavy rain, and compares them to the dashed line of the rain rate and the retained water curve for the `No gel' packing.  Three gel number ratios are tested: $\alpha =0.01$, $\alpha =0.02$, and $\alpha =0.04$.  Similar to the `No gel' curve, these curves follow the dashed line of the rain rate right after rain starts but deviate from it at later times.  As defined in Fig.~\ref{NoGel}, these time points refer to $t_c$ and illustrate how long rain water may take to penetrate through a dry packing and reach the water table.  As the $\alpha$ value in a packing increases, the packing shows a larger $t_c$.  This is reasonable since the hydrogel particles may lock part of the rain water inside by swelling and slow down the infiltration speed of the rest rain water.  When looking closely, we also notice another difference between the `No gel' curve and the other curves with hydrogel particle presenting ($\alpha>0$): the `No gel' curve approaches a quasi-stable value almost right after $t=t_c$; while the other curves show a long transition region starting at $t=t_c$ and ending when the increment on the curve becomes ignorable and a quasi-stable state approaches.  The appearance of the transition region is due to the continuously swelling of the hydrogel particles that distribute in the uniformly wetted top layer and in the fully-built water channel below it.  Thus the length and the shape of the transition region depend on multiple factors, including the mixing depth, the gel number ratio, and the gel swelling properties.  In these shallow mixing cases, since the mixing depth is comparable to the size of the uniform wetted region, the added dry hydrogel particles are all able to contact rain water in a short time after rain starts.  They absorb water, swell in size, perturb the soil pore structure around them, and expand the uniformly wetted top region of a packing vertically.  This process is time-consuming and the expansion of the wetted region can be clearly observed by eye.  For the three packings shown in Fig.~\ref{Mixture}~(a), both our observation and the obtained retained water curves indicate that they have approached a quasi-stable state after $3$~hours of raining ($t=180$~minutes) and a shorter time is required to reach quasi-stable as the the gel number ratio $\alpha$ decreases.  We compare the amount of the retained water at $t=180$~minutes and find that its increment is almost proportional to the gel number ratio $\alpha$.  We also estimate the expanding volume of a packing at the quasi-stable state and find that it is consistent with the total swelling size of the applied hydrogel particles in a packing.

When we extend the mixing depth to a larger value, for example to $10$~cm, the situation is a bit different.  Many added dry hydrogel particles are located in the water channel region and can not contact rain water at the early time of the rainfall.  If the hydrogel particles that luckily located in or near the water channel can significantly modify the water channel and lead water to them, they may get a chance to swell at a later time of the rainfall; if not, they will stay in dry for ever.  For this situation, it takes too much time to reach a quasi-stable state but the extended transition region usually shows a linear part whose slope closely relates to the gel number ratio $\alpha$ in a packing.  Fig.~\ref{Mixture}~(b) summarizes the retained water curves for packings with dry hydrogel particles uniformly mixed into top $10$~cm depth under a heavy rain.  Here four gel number ratios are tested: $\alpha =0.01$, $\alpha =0.02$, $\alpha =0.04$, and $\alpha =0.1$.  The first three values are the same as the shallow mixing cases shown in Fig.~\ref{Mixture}~(a) while the last value is applied to test if we can prevent the channel formation by rising the gel number ratio in a soil packing.  In the figure, we see that the retained water curves of the first three $\alpha$ values again show different delays on $t_c$ compared to the `No gel' curve.

\begin{figure}
\includegraphics[width=3in]{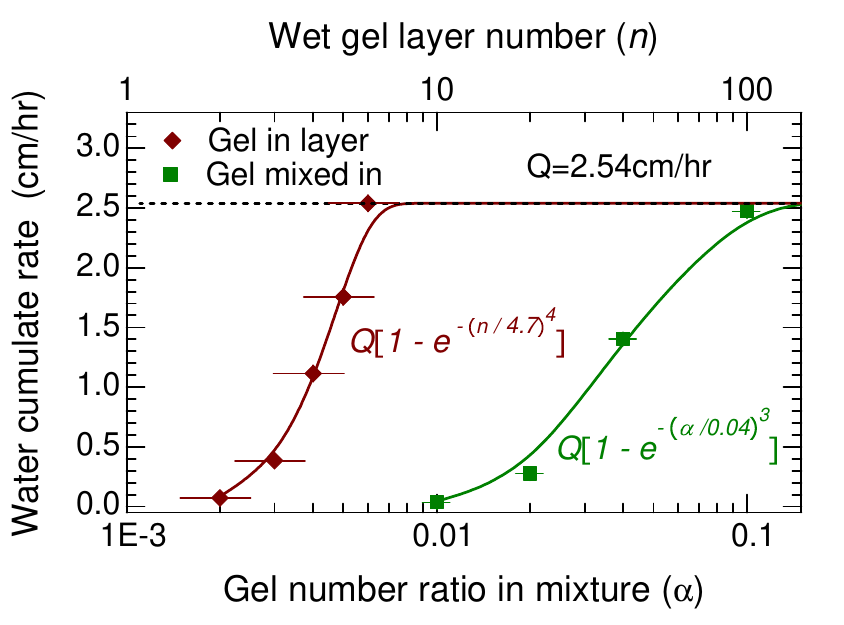}
\caption{(Color online)  Variation of the rain water accumulation rate with the mass of hydrogel particle additives in $1$~mm hydrophilic glass bead packing, with dry hydrogel particles ($0.3-0.5$~mm in axis) uniformly mixed in top $10$~cm region and placed in a layer under top $10$~cm soil, at a rain rate of $Q=2.54$~cm/hr.  The gel number ratio $\alpha$ is defined in Eq.~(\ref{GelNumRatio}); while the wet gel layer number $n$ is determined in Eq.~(\ref{GelLayer}).  The data points are obtained from the slopes of the linear fits in Fig.~\ref{Mixture}~(b) and in Fig.~\ref{LayerNum} respectively.  Dashed line marks the rain rate and the solid lines are sigmoidal fits of the data.  When the same amount of hydrogel particles are applied, the accumulation rate of rain water is far larger in packing with hydrogel particles placed in a layer under the ground than uniformly mixed into a top region of soils. }
 \label{CumulateRate}
\end{figure}

When compared with the shallow mixing cases shown in Fig.~\ref{Mixture}~(a), the transition regions are longer and have a more obvious linear part.  We fit the major parts of these transition regions to a line, determine the slopes, and plot the values against its gel number ratio $\alpha$ in Fig.~\ref{CumulateRate}.  These slopes have a unit of centimeters per hour, the same as the rain rate, and they tell us the accumulation rate of rain water in a packing.  Here we may consider them as a measure of the changes on the water channel and use the plot in Fig.~\ref{CumulateRate} to determine how gel number ratio affects the efficiency of hydrogel particles on modifying the water channel.  We also fit the quasi-stable region of the `No gel' curve to a line and confirm that its slope value is very close to zero and far smaller than those obtained from the mixture packing.  For packing with $\alpha =0.1$, the retained water curve follows the dashed line of the rain rate all the time and no deviation occurs even after $3$ hours of rain ($t=180$~minutes).  No rain water flows out the packing or ponds on soil surface -- the added hydrogel particles are sufficient to absorb all of the supplied water in time.  We again extract the slope of this curve and add its value to the plot in Fig.~\ref{CumulateRate}.

In Fig.~\ref{CumulateRate}, we fit the obtained slope values to a sigmoidal function given below:
\begin{equation}
   {\rm Water~accumulation~rate} = Q[1-e^{-(\alpha/0.04)^3}]  ~.
\label{Mixting}
\end{equation}
Here, rain rate $Q$ is the maximum water accumulation rate of the system and the fit gives us a critical gel number ratio of $\alpha_{cri}=0.04$.  More discussions on this figure will be given in the next section.

\begin{figure}
\includegraphics[width=3in]{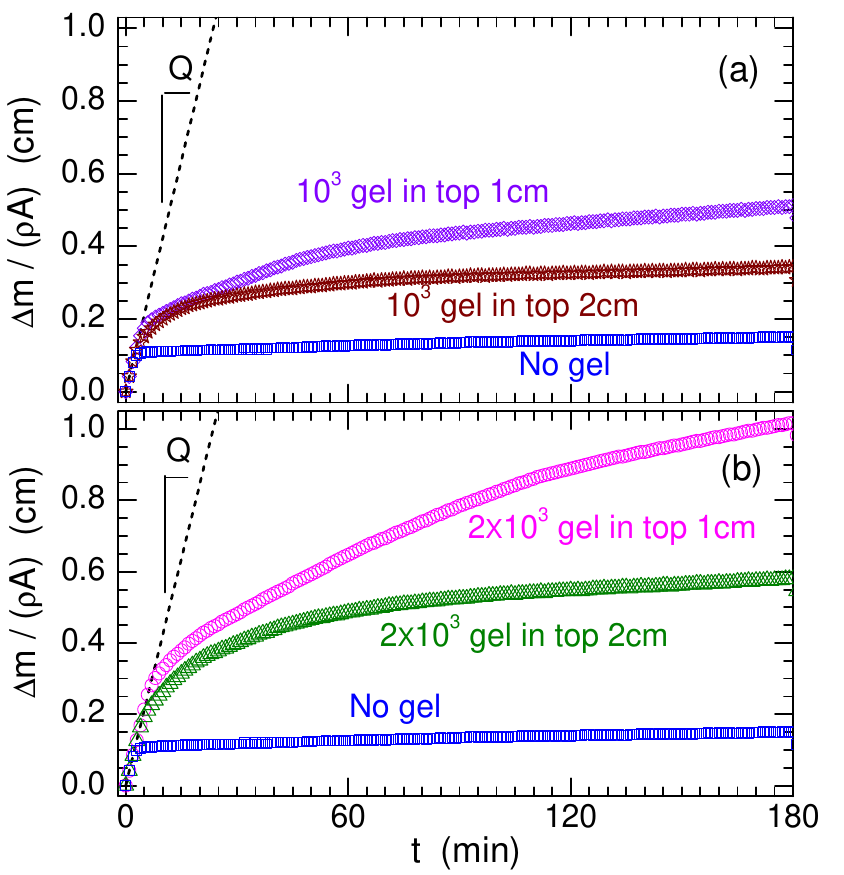}
\caption{(Color online)  Variation of the retained water with time for $1$~mm hydrophilic glass bead packing under a heavy rain, with (a) $10^3$ and (b) $2 \times 10^3$ dry hydrogel particles ($0.3-0.5$~mm in axis) uniformly mixed into top $1$~cm and top $2$~cm region of the soil packing respectively.  The initial packing height above the water table is $\Delta H=12.5$~cm for all packing.  The mass of the retained water $\Delta m$ is scaled by the water density $\rho$ and the cross-sectional area $A$ of the packing.  Rain starts at time $t=0$ with a rain rate of $Q=2.54$~cm/hr.  During the rain, the water table remains stable.  More rain water is retained when hydrogel particles are concentrated in a shallower top soil layer in the packing.  }
 \label{Concentrate}
\end{figure}

To further demonstrate the effect of mixing depth, we compare the retained water curves obtained from packings with the same number of hydrogel particles uniformly mixed into different deep top region under a heavy rain.  A thousand dry hydrogel particles are applied in Fig.~\ref{Concentrate}~(a); while two thousand particles are applied in Fig.~\ref{Concentrate}~(b).  The mixing depth is chosen to be $1$~cm and $2$~cm for each case.  In both Fig.~\ref{Concentrate}~(a) and (b), we see that more rain water is retained in soil packing with hydrogel particle additives concentrated to a shallower top layer.  The reason is that extending the mixture region decreases the gel number ratio in the mixture region thus reduces the number of the dry hydrogel particles that are able to contact water during the rain.  Without coupling rain water channelization phenomenon into the measurements, this influence will never be noticed.

\subsection{Drainage after rainfall}

\begin{figure}
\includegraphics[width=3in]{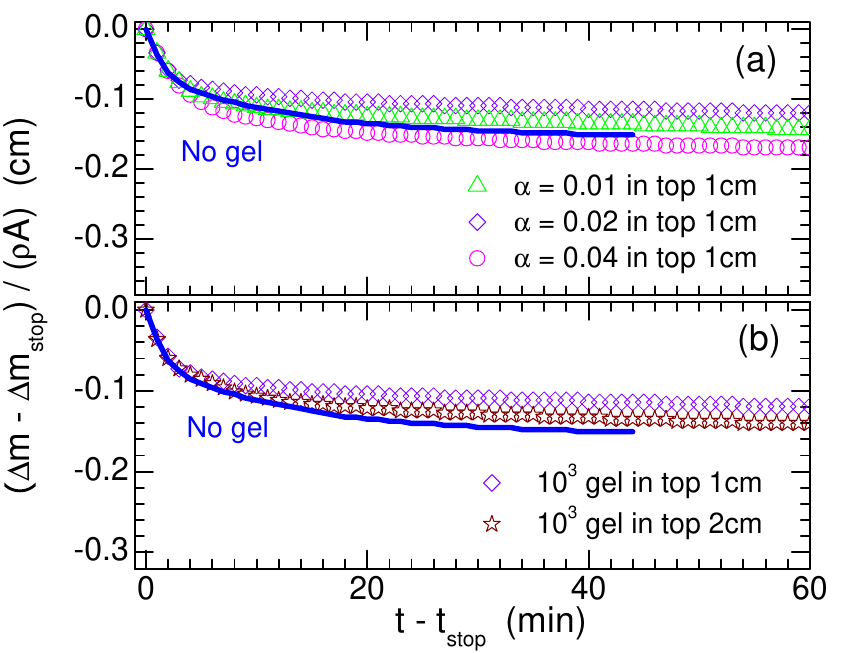}
\caption{(Color online)  Variation of the draining water with time for $1$~mm hydrophilic glass bead packing after rain stops, with dry hydrogel particles ($0.3-0.5$~mm in axis) uniformly mixed in top $1$~cm and top $2$~cm of the soil packing.  Results in (a) are obtained from packings shown in Fig.~\ref{Mixture}~(a); while results in (b) are obtained from packings shown in Fig.~\ref{Concentrate}~(a).  Rain stops at time $t=t_{stop}$, when the retained water in a packing is quasi-stable.  The mass of the retained water at that time is labeled as $\Delta m_{stop}$.  The water table in the soil packings is stable during the drainage.  The mass of the draining water ($\Delta m-\Delta m_{stop}$) is scaled by the water density $\rho$ and the cross-sectional area $A$ of the packing.  $\alpha$ is the number ratio of hydrogel particles to glass beads and is determined in Eq.~(\ref{GelNumRatio}).  Packings with hydrogel particles uniformly mixed in top region show similar drainage curves as the `No gel' one. }
 \label{MixtureDrain}
\end{figure}

Another way to determine if the swollen hydrogel particles significantly change the soil pore structure and it hydraulic conductively is to monitor the drainage behavior of packings after stopping the rain. As a continuation of the experiments shown in Fig.~\ref{Mixture}~(a), we stop the rain at $t=180$~minutes for all the packings, record their drainage curves, and compare with the `No gel' case.  In Fig.~\ref{MixtureDrain}~(a), we see that the drainage curves of the mixture packings behave the same as that obtained from the `No gel' packing.  Within the gel number ratio range we tested, the swollen hydrogel particles cannot efficiently slow down or reduce the gravitational drainage of the rain water.  We also stop the rain at $t=180$~minutes for packings shown in Fig.~\ref{Concentrate}~(a) and compare their drainage curves in Fig.~\ref{MixtureDrain}~(b).  Again, we see all the drainage curves collapse together.  No matter appearing in the top uniform infiltration layer or in the water channel, the swollen hydrogel particles do not affect the drainage of the retained rain water in sandy soil pores.  After free drainage, the final remaining water in a packing is mostly locked in the swollen hydrogel particles.  Comparing its value to the swelling ratio of hydrogel particles, we may estimate the percentage of the swollen hydrogel particles in a packing.  For examples, when a thousand dry hydrogel particles are applied in a model sandy soil packing, about $80$\% of them swells under rain when they are uniformly mixed in top $1$~cm region, but only $40$\% of them swells under rain when they are uniformly mixed in top $2$~cm region.

\section{Hydrogel particles placed in a layer under the ground }

Previous studies~\cite{Elshafei92, Elshafei94, Singh} have reported that when too many hydrogel particles are mixed into a shallow top layer of sandy soils they may swell to form a `gel shell' which prevents the infiltration of water into soil and thereby increases the surface water runoff.  We may build a similar structure under the ground to slow down rain water drainage and create extra water reservoirs near the plant roots, without runoff, as follows.  Instead of mixing, we place dry hydrogel particles in a layer under the surface of a model sandy soil packing with a height of $H=12.5$~cm.  We define a wet gel layer number $n$ to replace the gel number ratio $\alpha$ used in the mixing cases.  The wet gel layer number $n$ represents the maximum number of the pure wet gel layers that the added hydrogel particles can form in a packing with sufficient water supply.  It is estimated as
\begin{eqnarray}
   n &=& N_{gel}/ N'_{onelayer}  \\
     &=& \frac{M_{gel}/m_{gel}}{A/(\delta d)^2} ~.
\label{GelLayer}
\end{eqnarray}
Here, $N_{gel}$ is the total number of hydrogel particles that are placed under the ground; $N'_{onelayer}$ is the number of the hydrogel particles that are required to form a singe wet gel layer; $M_{gel}$ is the total mass of the added hydrogel particles and $m_{gel}$ is the average mass of a single hydrogel particle; $A$ is the cross-sectional area of the soil sample packing; $d=0.4\pm0.1$~mm is the average axis size of dry hydrogel particles; and $\delta=6$ is the free swelling ratio of hydrogel particles in sufficient DI water obtained by microscope measurements.

\begin{figure}
\includegraphics[width=3in]{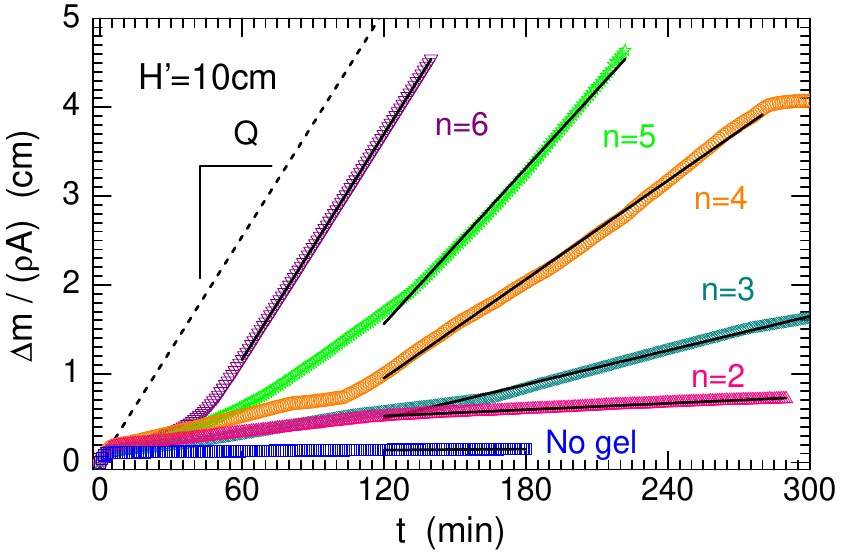}
\caption{(Color online)  Variation of the retained water with time for $1$~mm hydrophilic glass bead packing under a heavy rain, with different amount of hydrogel particles ($0.3-0.5$~mm in axis) placed in a layer in dry at a depth of $H'=10$~cm under the soil surface.  $n$ is defined in Eq.~(\ref{GelLayer}) and represents the maximum number of the wet gel layers the added hydrogel particles can build in a packing with sufficient water supply.  The initial packing height above water table is $\Delta H=12.5$~cm for all packing.  The mass of the retained water $\Delta m$ is scaled by the water density $\rho$ and the cross-sectional area $A$ of the packing.  Rain starts at time $t=0$ with a rain rate of $Q=2.54$~cm/hr.  During the rain, the water table remains stable.  Dashed line marks the rain rate; the solid lines are linear fits to determine the rain water cumulate rate in each packing.  The well-built wet gel layers partially clog the water channel and force part of the rain water to cumulate in pores of the model sand soil.  }
 \label{LayerNum}
\end{figure}

Fig.~\ref{LayerNum} summarizes the retained water curves obtained from packings with different mass of dry hydrogel particles placed in a layer at $H'=10$~cm depth under the soil surface.  The wet gel layer number in these packings varies from $n=2$ to $n=6$.  Rain starts at $t=0$ and the rain rate $Q=2.54$~cm/hr is marked by the dashed line.  From the figure, the first thing we notice is that within the $n$ range we tested the dry gel layer under the ground cannot efficiently prevent the full formation of the water channel: all the retained water curves show a deviation from the dashed line of rain rate at an early time of the rainfall, the same as the `No gel' case shown by the blue square symbol.  Their $t_c$ values are slightly larger than that of the `No gel' case and close to those obtained from the uniformly mixing cases shown in Fig.~\ref{Mixture}.  This is reasonable since the hydrogel particles we applied here are not sufficient to from a dense layer in dry and their swelling is time-consuming.  Another obvious thing we find is that the transition region in these curves significantly differs from that of the mixing cases, which shows a single linear increase until reaching the quasi-steady state.  The transition regions shown in Fig.~\ref{LayerNum} are more complex and most of them can be clearly separated into two parts: a slowly increasing region corresponds to the formation of the wet gel layers; while a sharp linear increasing region corresponds to the accumulation of the rain water in the sandy soil due to the clogging effects of well-built wet `gel shell' under the ground.  As the wet gel layer number $n$ increases, less time is consumed to full build the wet gel layers and more rain water is retained inside the packing.  The reason is that when the density of the dry hydrogel particles increases in a packing more hydrogel particles locate in the cross-section of the water channel and their swelling horizontally directs more rain water to their dry neighbors.  During the experiments, through the transparent sample column we can clearly see an optically-clear swollen hydrogel region built across the soil packing and the model sandy soil above is lifted up.  The size of the swollen gel region is consistent with that estimated from the wet gel layer number $n$ in the packing.

The fully-built wet gel layers can be very efficient in clogging the rain water.  In Fig.~\ref{LayerNum}, we see that the second part of the transition region in the different retained water curves all have very different slopes.   Experimentally, we observed a fully saturated region in the model sandy soil forming and growing upward right above the optical-clear swollen hydrogel region.  The clogging efficiency depends on the wet gel layer number $n$: the saturated region in packings with high $n$ values keeps growing up to the top of the packing and then ponding on the soil surface; the saturated region in packings with low $n$ values usually stops growing up at certain level and reaches a quasi-steady state.  We stop the rain right before water begins to pond on soil surface for the first case and when system reaches the quasi-steady state for the second case. Since the major rain water storage mechanism for packings shown in Fig.~\ref{LayerNum} is to create water reservoir in sandy soil itself rather than to lock water inside hydrogel particle additives, we fit the second part of the transition regions to a solid line, as shown in the figure, use the slope values to represent the rain water accumulation rates in these soil packings, and compare to the uniformly mixing cases discussed in the last section.  The variation of the slopes with the corresponding wet gel layer number $n$ are also shown in Fig.~\ref{CumulateRate} and it behaves very similar to that obtained from the mixing cases but shifts to left in x-axis.  We again fit the data using the same sigmoidal function and have
\begin{equation}
   {\rm Water~accumulation~rate} = Q[1-e^{-(n/4.7)^4}]  ~.
\label{Layer}
\end{equation}
Here, rain rate $Q$ is the maximum water accumulation rate of the system and the fit gives us a critical wet gel layer number of $n_{cri}\approx5$.  If we convert both $n_{cri}$ and $\alpha_{cri}$ back to the mass of the added hydrogel particles, we will see that the former one is far smaller than the later one.  So we come to the conclusion that placing hydrogel particle additives in a layer under the ground is more efficiency in improving sandy soil water storage than uniformly mixing them into soils.  The reason is because by placing hydrogel particles in a layer under the ground we successfully add a new water storage mechanism into the system.  Beside storing rain water inside the hydrogel particles, the clogging effects of swollen hydrogel particles is also carefully applied to create a temporary water reservoir inside sandy soils, which largely accumulates rain water and extends the soil region where rain water can reach.

\subsection{Drainage after rainfall}

\begin{figure}
\includegraphics[width=3in]{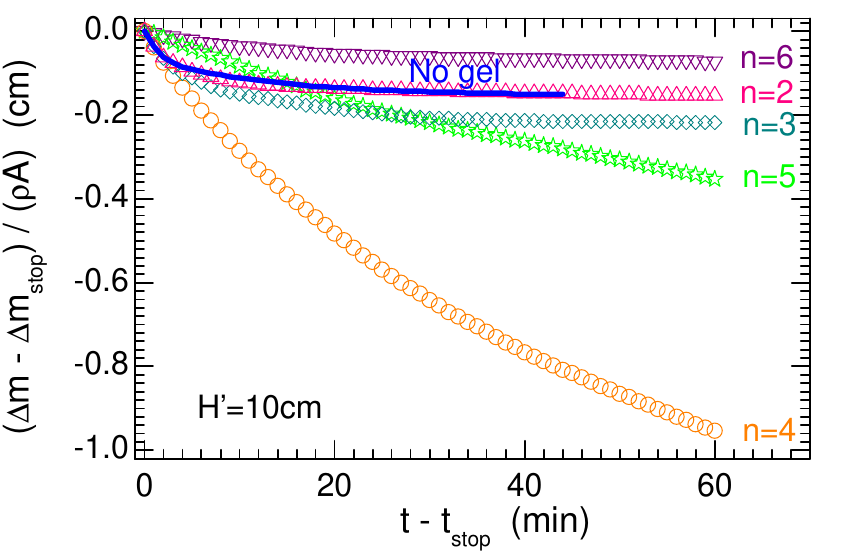}
\caption{(Color online)  Variation of the draining water with time for $1$~mm hydrophilic glass bead packing after rain stops, with different amount of hydrogel particles ($0.3-0.5$~mm in axis) placed in a layer in dry at a depth of $H'=10$~cm under packing surface.  These results are obtained from the same packings shown in Fig.~\ref{LayerNum}.  $n$ is defined in Eq.~(\ref{GelLayer}) and represents the maximum number of the wet gel layers the added hydrogel particles can build in a packing with sufficient water supply.  Rain stops at $t=t_{stop}$ when a packing is in quasi-steady state ($n\leq4$) or when a packing is fully saturated to its surface by rain water ($n>4$).  The mass of the retained water at $t_{stop}$ is labeled as $\Delta m_{stop}$.  During the drainage, the water table remains stable.  The mass of the draining water ($\Delta m-\Delta m_{stop}$) is scaled by the water density $\rho$ and the cross-sectional area $A$ of the packing.  The maximum drainage happens at the $n=4$ packing and the drainage decreases for packings with smaller $n$ values and with larger $n$ values.  }
\label{LayerNumDrain}
\end{figure}

The drainage behavior for packings with hydrogel particles placed in a layer is far more complex and interesting than that of the uniformly mixing cases.  Fig.~\ref{LayerNumDrain} collects the drainage curves obtained from exactly the same soil packings shown in Fig.~\ref{LayerNum}.  Differing from the uniformly mixing cases, the rain stop time $t_{stop}$ for these drainage curves varies from each other.  As we mentioned before, the strong clogging effects due to the well-built wet gel layer may prevent a soil packing to reach a steady state under the rainfall.  So, for $n=6$ and $n=5$ packing, we have to stop the rain when its fully saturated region grows to the soil surface to prevent water ponding.  For $n=4$, $n=3$, $n=2$ packings, we again stop rain when they reach the quasi-steady state.  The values of $t_{stop}$ for these packings can be determined from the last data point in each curve shown in Fig.~\ref{LayerNum} and the maximum $t_{stop}$ value is $300$~minutes.  In Fig.~\ref{LayerNumDrain}, we see that most of the drainage curves obtained from packings with hydrogel particles place in a layer under the ground show large deviation from the `No gel' curve.  An interesting thing we notice is that these curves even not changes monotonously.  The maximum drainage occurs in the $n=4$ packing.  For $n<4$ packings and $n>4$ packings, the drainage reduces as $n$ value decreases or increases, respectively.  Considering the difference of the drainable water at $t=t_{stop}$ in each packing, we should not be surprised by the results.  The drainable water refers to the amount of water that only temporarily stays in a packing after rain stops and does not lock by the hydrogel particles or by the capillary forces in soil pores.  Due to the well-built wet gel layers, there is a dramatic increase in the drainable water from the $n=2$ packing to the $n=4$ packing.  Therefore, for the $n\leq4$ packings the drainage behavior is dominated by the amount of drainable water in a packing at $t=t_{stop}$; while for the $n\geq4$ packings the drainage behavior is mainly determined by the clogging efficiency of the well-built wet gel layers on reducing water draining speed.

\subsection{Depth of gel particle layer}

\begin{figure}
\includegraphics[width=2.75in]{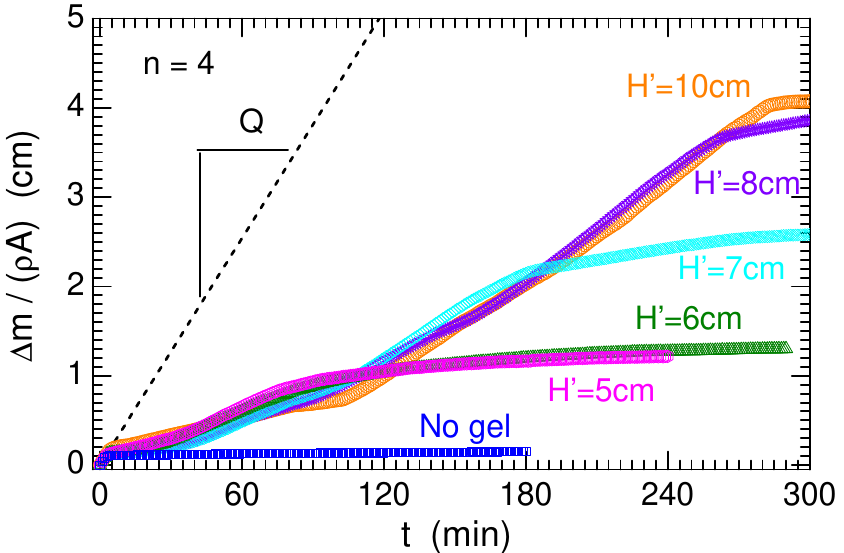}
\caption{(Color online)  Variation of the retained water with time for $1$~mm hydrophilic glass bead packing under a heavy rain, with two thousand hydrogel particles placed in a layer in dry at different depths $H'$ under the soil surface.  The initial packing height is $\Delta H=12.5$~cm for all packing.  The mass of the retained water $\Delta m$ is scaled by the water density $\rho$ and the cross-sectional area $A$ of the packing.  Rain starts at $t=0$ with a rain rate of $Q=2.54$~cm/hr.  During the rain, the water table remains stable.  Roughly the same amount of time is taken to build the wet gel layers but their efficiency in showing down the rain water drainage is very different.  Only the wet gel layers formed under sufficient depth under ground are able to significantly reduce the rain water penetration speed and force part of the rain water to cumulate in the pores of the model sandy soil about them. }
 \label{LayerDepth}
\end{figure}

Another important parameter that may affect the clogging efficiency is the location of the wet gel layers.  Since the soil confinement on the wet gel layers comes from the weight of the model sandy soil above them, placing hydrogel particles in a shallower location under ground reduces the soil confinement thus may modify the efficiency of the wet gel layers in slowing down rain water drainage.  Fig.~\ref{LayerDepth} shows the retained water curves obtained from packings with the same amount of hydrogel particles placed in a layer in dry at different depths $H'$ under the soil surface.  The base packing (`No gel' one) has a packing height of $\Delta H=12.5$~cm.  Two thousand hydrogel particles, giving a wet gel layer number of $n=4$, are added in each packing but the locations varies from $H'=5$~cm to $H'=10$~cm.  Rain starts at $t=0$ with a rain rate of $Q=2.54$~cm/hr, as marked by the dashed line in the figure.  In Fig.~\ref{LayerDepth}, we notice several things: first, changing the location of the gel layer does not affect the formation time of the water channel and the $t_c$ values for all the packings are roughly the same; second, packings with hydrogel particles placed in a shallower location (smaller $H'$) require less time to reach the quasi-steady state and retain less rain water at that state; third, the slopes of the transition regions for all these curves are very close to each other.  In detail, when hydrogel particles are placed in a shallow location in the packing, the corresponding retained water curve no longer shows a clear sharp change in the slope of the transition region, such as $H'=5$~cm and $H'=6$~cm curves.  The reason is that hydrogel particles feel less vertical confinement in a shallow location of the packing.  They can swell more freely and pack more loosely under the rain, which strongly reduces their water clogging efficiency.  During the experiments, we indeed observed the following facts: the thickness of the wet optical-clear hydrogel region formed in $H'=10$~cm packing is only about $70$~\% of that formed in $H'=5$~cm packing; the size of the fully-saturated region created above the wet gel layers is less than $0.5$~cm in $H'=5$~cm packing but increase to about $3$~cm in $H'=7$~cm packing and $8$~cm in $H'=10$~cm.  Fig.~\ref{LayerDepth} clearly demonstrates that the vertical soil confinement plays a critical role in building the wet gel layers with efficient clogging effects.

\begin{figure}
\includegraphics[width=2.75in]{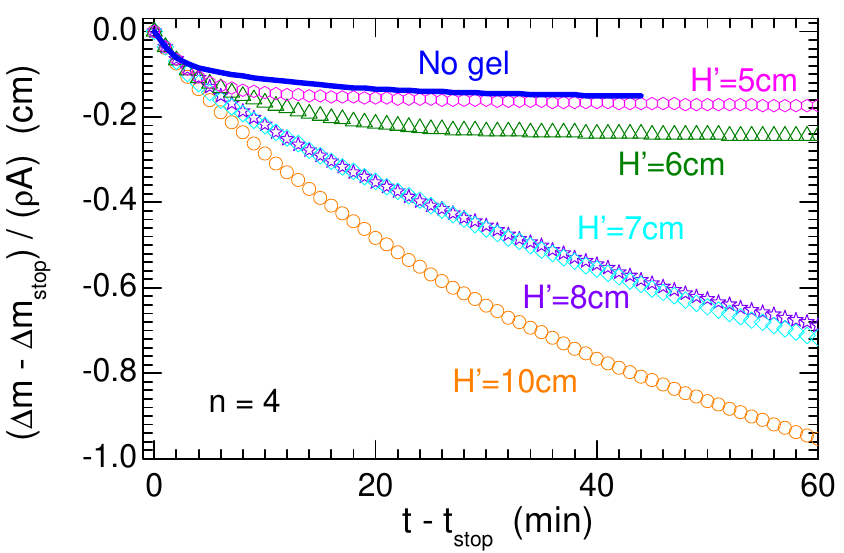}
\caption{(Color online)  Variation of the draining water with time for $1$~mm hydrophilic glass bead packing after rain stops, with two thousand hydrogel particles placed in a layer in dry at different depths ($H'$) under the soil surface.  These results are obtained from the same packings shown in Fig.~\ref{LayerDepth}.  Rain stops at $t=t_{stop}$ when a packing reaches its quasi-steady state.  The mass of the retained water at $t=t_{stop}$ is labeled as $\Delta m_{stop}$.  During the drainage, the water table remains stable.  The mass of the drainage water ($\Delta m-\Delta m_{stop}$) is scaled by the water density $\rho$ and the cross-sectional area $A$ of the packing.  More water drains out when hydrogel particles are placed in deeper location under the soil surface. }
 \label{LayerDepthDrain}
\end{figure}

We also compare the drainage behavior for packings with the same gel layers formed at different depth under the ground after rain stops.  Fig.~\ref{LayerDepthDrain} collects the drainage curves for exactly the same packings shown in Fig.~\ref{LayerDepth}.  Again, we stop the rain when a packing reaches quasi-steady state.  The $t_{stop}$ values vary for these packings and can be obtained form the last data point of the curves shown in Fig.~\ref{LayerDepth}.  In Fig.~\ref{LayerDepthDrain}, we see that most of the drainage curves significantly deviate from the `No gel' case.  And this time a monotonously change in the drainage curves is seen in the figure as the $H'$ value decreases: the draining water decreases as the wet gel layers locates closer and closer to the soil surface.  The main reason is that the amount of drainable water stored in packings with large $H'$ values at $t=t_{stop}$ is far larger than that stored in packings with small $H'$ values.  Comparing the amount of water that retains in a packing at $t_{stop}$ and that drains out after rain stops, we find that even after hours of free draining, there is still more water remaining in packings when larger $H'$ value is applied.

\section{Conclusion}

In conclusion, we built a $3$D laboratory set-up to study how rain water is transported and stored in a model sandy soil when hydrogel particles are applied into the soil by different methods.  The set-up mimics a heavy rain condition in nature and measures the mass of the retained water in a soil packing during the rainfall and after rain stops.  The sample column we used is transparent and allow us to also observe certain experimental phenomena by eye.  The data and the visual observations are combined to determine the principle mechanisms that control the transport and storage of rain water in sand soils with commercial superabsorbent additives.

The model sandy soil we chosen has a well-known pore structure and very hydrophilic grain surfaces.  It shows extremely poor water-holding capacity under rain -- that amount of the retained water in a pure soil packing is $30$ times less than what is required to wet the whole packing.  Our study shows that rain water only uniformly wets a shallow top layer of the soil packing and then an instability occur on the wetting front and grow into a fingered flows that finally penetrates through the packing and forms a narrow water channel.  Since the saturated hydraulic conductivity of the model sandy soil is far larger than the rain rate, both the wet top layer and the water channel in the soil packing maintain a low saturation level during the rainfall.  Once the water channel is fully formed, rain water follows this path to drain out and the rest of the soil remains dry.

When dry hydrogel particles are uniformly mixed into the model sandy soil, under the same raining condition the amount of the retained water in the packing increases and the increment strongly depends on the mixing depths and the gel number ratio in the mixture region.  Not all the hydrogel particle additives can have contact with the rain water; only those lucky ones located in the wet top layer or in the water channel are able to absorb water.  Within the gel number ratio range we tested, the presence of hydrogel particle additives cannot prevent the full formation of the water channel but only delay it a bit in time.  The reason is that their water supply is limited and their swelling is time-consuming.  The swelling of these hydrogel particles slowly modifies the size and the shape of the water channel so that rain water may be directed to flow through their dry neighbors.  Here the major mechanism that enhances the soil water retention is to lock as much rain water as possible into the hydrogel particles rather than increase the capillary storage of water in sandy soil by modifying soil pore structure.  Our study on the drainage behavior of these packings further confirms this conclusion.  Therefore, the key of optimizing the method is to maximize the percentage of added hydrogel particles that can contact and absorb rain water by carefully choosing the mixing depths and the gel number ratio in the soil.

When dry hydrogel particles are placed in a layer deeply under ground, in the same raining condition the amount of the retained water in the packing increases even more significantly than that in the uniformly mixing cases.  The hydrogel particles applied by this method neither prevent the fingered flow to penetrate nor strongly slow down the flow speed in the water channel until the optic-clear wet gel layers are well-built across the soil packing under the rainfall.  Beside locking rain water in hydrogel particles, a new rain water storage mechanism is induced here using the clogging effects of the wet gel layers to enhance the capillary storage of water in sandy soil pores.  The building of wet gel layers is time-consuming and their efficiency largely depends on the forming location and the number of the wet gel layers.  Wet gel layers formed deeply under ground are very efficient in clogging rain water thus creating large temporary water reservoirs in sandy soils above them; wet gel layers formed in a shallow location are not so efficient in clogging rain water but may absorb more water inside the gel layers.  For wet gel layers formed at the same depth, increasing the number of the wet gel layers enhances the clogging effects and the water-holding capacity in a packing.  We also monitor the drainage behaviors in these cases after rain stops and find that it is far more complex than that obtained in the mixing ones.  The reason is that the drainage behavior here is not only controlled by the efficiency of the wet gel layers in slowing down draining speed but also decided by the total drainable rain water retained in each packing at the time rain stops.

Our study has elucidated many interesting features in rain water transport and storage in a sandy soils with and without superabsorbent additives.  And it shows how the additives may be efficiently used, in light of the formation of fingered flows.  Furthermore it brings new questions into focus: how exactly does the presence of the hydrogel particles modify the fingered flowing paths of rain water in sandy soils?  How does particle swelling disturbs the pore structure of sandy soils?  If there is more than one channel, what sets the separation length?  To help answer such questions, a quasi two-dimensional version of our set-up has been constructed in order to to directly visualize channel morphology and kinetics \cite{Wei2D, Cejas2D}.

\begin{acknowledgments}
We thank Jean-Christophe Castaing, Zhiyun Chen, Cesare M. Cejas, and Remi Dreyfus for helpful discussions. We also thank Degussa Inc. for kindly providing us the hydrogel particle samples.  This work is supported by the National Science Foundation through Grants MRSEC/DMR-112090 and DMR-1305199.
\end{acknowledgments}


\bibliography{Rain3D_References}


%
\end{document}